\documentclass[aps,prl,reprint,superscriptaddress,nofootinbib,floatfix]{revtex4-2}

\usepackage{amsmath,amssymb,bm}
\usepackage{booktabs}
\usepackage{braket}
\usepackage{graphicx}
\usepackage{color}

\DeclareMathOperator{\Tr}{Tr}

\begin{document}

\title{Symmetry-resolved parent Hamiltonians for entangled bosonic cat resources}

\author{\v{S}imon Br\"{a}uer}
\author{Klaus M{\o}lmer}
\affiliation{Niels Bohr Institute, University of Copenhagen, Jagtvej 155 A, DK-2200 Copenhagen, Denmark}
\date{\today}

    \begin{abstract}
       Schr\"odinger cat states, formed from superpositions of well-separated coherent states, are promising resources for bosonic quantum information. We construct parent Hamiltonians for entangled multimode cat states by separating the selection of local coherent-state branches from the correlations and symmetries that determine their global superposition. GHZ-, cluster-, and W-type cat states illustrate three distinct mechanisms for this construction. In the limit of well-separated branches, the resulting oscillator Hamiltonians reduce to familiar qubit parent models based on correlations, stabilizers, and exchange symmetry. The same constraint structure also provides a common basis for coherent preparation, dissipative stabilization, and state certification.
    \end{abstract}

\maketitle

    \paragraph{Introduction---}\label{sec:introduction}The controlled preparation of nonclassical bosonic states is central to quantum information processing with oscillators. {Schr\"odinger cat, superpositions of separate coherent states $\ket{\pm\alpha}$,} form the basis of bosonic cat codes and provide a hardware-efficient route to protected logical encoding \cite{Hoshi2025,Mirrahimi2014, Ofek2016, Grimm2020Aug, Guillaud2019,Chamberland2022}. Their realization in superconducting and trapped-ion platforms has enabled logical control, noise-biased encoding, and quantum error correction \cite{Lescanne2020May,Fluhmann2019Feb,Cai2021}. An important next step is to extend these capabilities from individual oscillators to entangled bosonic resources distributed across multiple
    modes.
    
    Such multimode resources are relevant for distributed encodings, modular quantum architectures, and graph-based quantum information processing \cite{Wang2016,Chou2018}. Representative examples include GHZ-, W-, and cluster-type {cat} states, which realize qualitatively different forms of multipartite entanglement \cite{Greenberger1989,Dur2000,Zhu2024,Raussendorf2001, Pfister2019,Menicucci2006,An2009}. Their preparation, stabilization, and certification, however, are usually treated as separate and largely state-specific tasks. Driven and dissipative protocols can stabilize particular target states \cite{Leghtas2013,Xu2023Aug}, but they do not by themselves provide a general operator-level prescription for identifying which
    constraints uniquely define a desired multimode bosonic resource \cite{Albert2018}.
    
    A parent Hamiltonian provides such a description through the common kernel of a set of algebraic constraints. This viewpoint is useful beyond the formal identification of a ground state: the same constraint structure can guide coherent preparation through Hamiltonian engineering, dissipative stabilization through jump operators with the corresponding dark-state kernel, and certification through expectation values of the positive-semidefinite parent operator. Preparation, stabilization, and certification can therefore be organized around the same defining kernel rather than formulated as unrelated problems.
    
    The single-mode precursor of this idea appeared in the notion of catability, where positive-semidefinite oscillator operators were introduced to characterize coherent-state superpositions \cite{Brauer2026,Matulik2026Jun}. A branch constraint fixes the allowed coherent amplitudes, while an additional symmetry constraint selects the desired superposition. Although originally introduced for state evaluation, the zero-energy structure of these operators suggests a broader principle: compact oscillator Hamiltonians can encode the defining constraints of nonclassical bosonic states.
    
    Here we extend this principle to entangled multimode resources. A universal two-photon branch Hamiltonian first confines every mode to the coherent-state manifold $\{\ket{\alpha},\ket{-\alpha}\}$. Additional positive-semidefinite correlation or symmetry constraints then remove unwanted directions within the resulting branch manifold. For graph-state resources, the construction extends whenever these bosonic constraints reproduce a complete set of graph stabilizers inside the selected manifold.
    
    We demonstrate the framework using GHZ-, cluster-, and W-type cat states, which illustrate three distinct selection mechanisms: branch alignment and global symmetry, graph-dependent stabilizer selection, and fixed-defect selection followed by permutation symmetrization. For well-separated coherent branches, the resulting oscillator Hamiltonians reduce to stabilizer or exchange parent Hamiltonians  acting on encoded qubits \cite{Gottesman1997, Terhal2020, Kuchar2026Apr}. The construction therefore connects coherent-state bosonic engineering with discrete-variable parent-Hamiltonian methods and provides a common operator language for the preparation, stabilization, and certification of entangled bosonic resources.


    \paragraph{Constraint-based bosonic parent Hamiltonians---}\label{sec:framework} The construction is motivated by the  single-mode catability operator introduced in Refs.~\cite{Brauer2026,Matulik2026Jun}. For one bosonic mode with annihilation operator $a$, define 
        \begin{equation}
            H_{\mathrm{cat}}^{(\pm)} = \bigl(a^2-\alpha^2\bigr)^\dagger\bigl(a^2-\alpha^2\bigr) + \gamma\bigl(\mathbb I\mp\Pi\bigr),
            \qquad \Pi=(-1)^{a^\dagger a},
            \label{eq:single_cat}
        \end{equation}
    where $\gamma>0$.
    The first term is a branch constraint: since
        \begin{equation}
            a^2\ket{\pm\alpha}=\alpha^2\ket{\pm\alpha},
        \end{equation}
    its kernel is the two-dimensional coherent-state manifold
        \begin{equation}
            \mathcal B_\alpha = \operatorname{span} \bigl\{\ket{\alpha},\ket{-\alpha}\bigr\}.
            \label{eq:single_branch_manifold}
        \end{equation}
    The parity term resolves the remaining degeneracy and selects the even or odd cat state
        \begin{equation}
            \ket{\mathcal C_\pm} = \mathcal N_\pm \bigl(\ket{\alpha}\pm\ket{-\alpha}\bigr)
        \end{equation}
    as the unique zero-energy state. The construction therefore separates two logically distinct tasks: first selecting the allowed coherent branches and then selecting a specific superposition within their span.
    
    For $M$ bosonic modes with annihilation operators $a_j$, the universal branch Hamiltonian is
        \begin{equation}
            H_{\mathrm{br}} = \sum_{j=1}^{M} \bigl(a_j^2-\alpha^2\bigr)^\dagger \bigl(a_j^2-\alpha^2\bigr).
            \label{eq:multimode_branch}
        \end{equation}
    Its zero-energy manifold is $ \mathcal B_\alpha^{(M)} = \operatorname{span} \left\{ \ket{\bm s} = \ket{s_1\alpha,\ldots,s_M\alpha}:s_j=\pm1 \right\}, \dim\mathcal B_\alpha^{(M)}=2^M$.
    
    {The branch Hamiltonian fixes the local coherent-state support but not the correlations or superposition amplitudes between branch configurations.}
    
    {For a target state $\ket{\psi_0}\in\mathcal B_\alpha^{(M)}$, we introduce state-dependent positive-semidefinite constraints $H_\mu\geq0$ chosen such that $H_\mu\ket{\psi_0}=0$. 
        \begin{equation}
            H_{\mathrm{target}} = H_{\mathrm{br}} + \sum_{\mu}\lambda_\mu H_\mu, 
            \qquad \lambda_\mu>0.
            \label{eq:general_parent}
        \end{equation}}  
    Because all contributions are positive semidefinite, the zero-energy space is the common kernel
        \begin{equation}
            \ker H_{\mathrm{target}} = \mathcal B_\alpha^{(M)} \cap \bigcap_\mu \ker H_\mu.
            \label{eq:kernel_intersection}
        \end{equation}
    Whenever this intersection is one-dimensional, $H_{\mathrm{target}}$ is a frustration-free parent Hamiltonian whose unique ground state is the desired bosonic resource.
    
    {The additional constraints may enforce branch correlations, symmetries, graph-dependent relations, or coherent mixing within a selected sector.} A bosonic graph-state construction follows the same principle: the branch term fixes the local coherent support, while a complete set of graph-dependent constraints selects a one-dimensional common kernel inside $\mathcal B_\alpha^{(M)}$. The GHZ-, cluster-, and W-type states considered below illustrate three
    different realizations of this general mechanism.
    
    The common-kernel formulation also anticipates the certification role of the parent Hamiltonian. When the common kernel is one-dimensional, $ H_{\mathrm{target}}\geq0$ vanishes only on the desired state, and its expectation value quantifies violations of the defining constraints. Most local terms considered here are polynomial in oscillator operators and can be evaluated from suitable field moments, whereas {the swap contribution is accessible through mode mixing followed by parity detection}. The same parent operator can therefore specify the target state and provide the basis for energy- and moment-based certification witnesses \cite{Marek2024May,Brauer2025Aug, Kala2025May, Brauer2026, Provaznik2026Mar, Racz2026Jun}.
    

    \paragraph{Representative multimode cat resources---}\label{sec:multimode_examples}We now apply the common-kernel construction to representative multimode bosonic resources. The GHZ-, cluster-, and W-type cat states illustrate three distinct mechanisms for resolving the degenerate branch manifold: global branch alignment followed by  symmetry selection, graph-dependent correlation constraints, and fixed-defect selection followed by permutation symmetrization, respectively.

        \emph{GHZ-type cat states.-}\label{sec:ghz_bosonic}The $M$-mode GHZ-type cat states are coherent superpositions of the two globally aligned branch configurations,
            \begin{equation}
                \ket{\mathrm{GHZ}_{\pm}} = \mathcal N_{\mathrm{GHZ},\pm}  \left( \ket{\alpha}^{\otimes M} \pm \ket{-\alpha}^{\otimes M} \right),
            \label{eq:ghz_state}
            \end{equation}
        where $\mathcal N_{\mathrm{GHZ},\pm}$ is a normalization factor. Starting from the universal branch Hamiltonian in Eq.~\eqref{eq:multimode_branch}, the target state is selected in two steps. The first imposes correlations between the coherent branches of different modes, while the second fixes the symmetry of their remaining superposition.
        
        To select the globally aligned branch configurations, we introduce the positive-semidefinite branch-alignment Hamiltonian
            \begin{equation}
                H_{\mathrm{align}}^{(\mathrm{GHZ})} = \sum_{(i,j)\in\mathcal E} \lambda_{ij} (a_i-a_j)^\dagger(a_i-a_j),
                \qquad \lambda_{ij}>0,
            \label{eq:ghz_alignment}
            \end{equation}
        where $\mathcal E$ is a connected set of pairs of oscillator modes.
        For a branch configuration $\ket{\bm s}=\ket{s_1\alpha,\ldots,s_M\alpha}$, $(a_i-a_j)\ket{\bm s} = \alpha(s_i-s_j)\ket{\bm s}$.
        Consequently, every pair with $s_i\neq s_j$ acquires a positive energy. Since the set $\mathcal E$ connects all modes, the common
        zero-energy condition requires $s_1=s_2=\cdots=s_M$. The original $2^M$-dimensional branch manifold is therefore reduced to $ \mathcal B_{\mathrm{align}} = \operatorname{span} \left\{ \ket{\alpha}^{\otimes M},\ket{-\alpha}^{\otimes M} \right\}$. The remaining twofold degeneracy is resolved by the global photon-number parity
            \begin{equation}
                \Pi_G = (-1)^{\sum_{j=1}^{M}a_j^\dagger a_j} = \bigotimes_{j=1}^{M}\Pi_j ,
            \label{eq:global_parity}
            \end{equation}
        which exchanges the two aligned branches and satisfies
            \begin{equation}
                \Pi_G\ket{\mathrm{GHZ}_{\pm}} = \pm\ket{\mathrm{GHZ}_{\pm}}.
            \end{equation}
        The corresponding positive-semidefinite symmetry selector is
            \begin{equation}
                H_{\mathrm{par}}^{(\mathrm{GHZ},\pm)} = \gamma\left(\mathbb I\mp\Pi_G\right),
                \qquad \gamma>0.
            \label{eq:ghz_parity}
            \end{equation}

        The full parent Hamiltonian is therefore
            \begin{equation}
                H_{\mathrm{GHZ},\pm} = H_{\mathrm{br}} + H_{\mathrm{align}}^{(\mathrm{GHZ})} + H_{\mathrm{par}}^{(\mathrm{GHZ},\pm)}.
            \label{eq:ghz_parent}
            \end{equation}
        Its common kernel is one-dimensional and is spanned by $\ket{\mathrm{GHZ}_{\pm}}$, with the successive constraints reducing the zero-energy manifold according to $2^M \longrightarrow 2 \longrightarrow 1$. The $M=2$ realization, for which the reduction becomes $4\rightarrow2\rightarrow1$, is illustrated in         Fig.~\ref{fig:ghz_parent}; a complete uniqueness proof is given in the Supplemental Material~\cite{SM}.

            \begin{figure}[t]
                \centering
                \includegraphics[width=\linewidth]{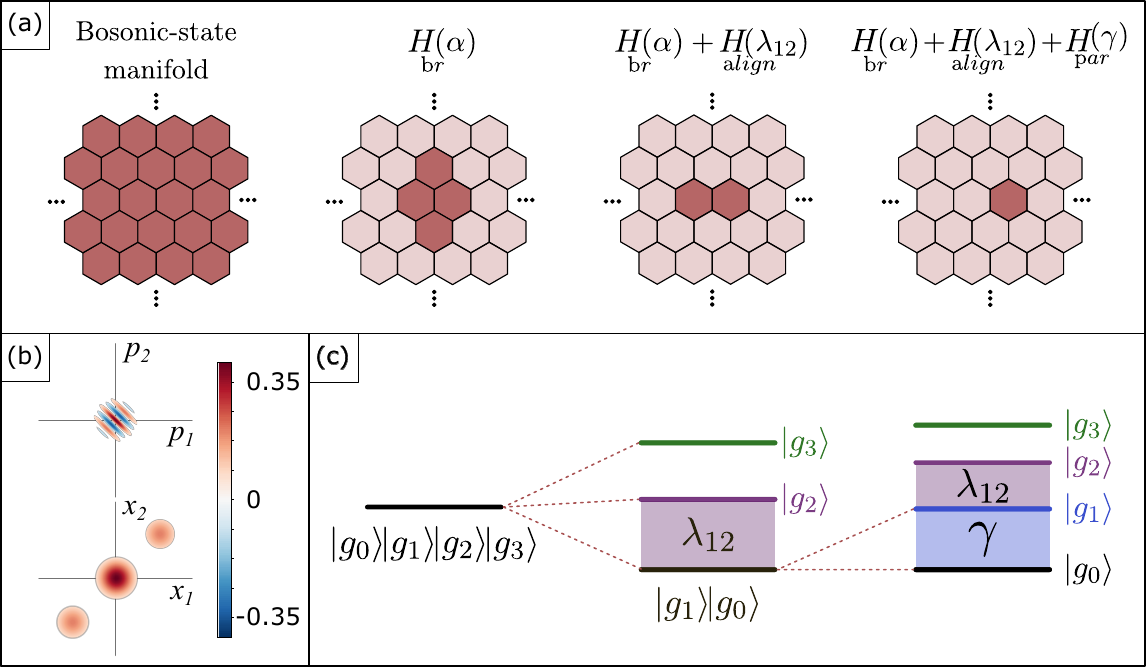}
                \caption{
                Constraint hierarchy for a two-mode GHZ-type cat state.
                (a) The branch Hamiltonian selects four coherent {state} configurations, the alignment constraint retains the two globally aligned branches, and global parity selects a unique GHZ-type superposition.
                (b) The slices $W(x_1,0,x_2,0)$ and $W(0,p_1,0,p_2)$ display the correlated branches and their interference, respectively.
                (c) Successive constraints reduce the zero-energy manifold from four dimensions to two and finally to one.
                }
                \label{fig:ghz_parent}
            \end{figure}

        { Any connected choice of $\mathcal E$ leaves the selected zero-energy state unchanged, while the couplings $\lambda_{ij}$ determine the excitation spectrum and branch-defect energies.
}
        \emph{Cluster-type cat states.-}\label{sec:cluster_bosonic}Unlike the GHZ construction, a cluster-type cat retains several coherent branch configurations and selects the target through graph-dependent correlations between them. {We illustrate this mechanism with a four-mode target state
            \begin{equation}
                \ket{C} = \frac{1}{2} \left( \ket{+,+} + \ket{+,-} + \ket{-,+} - \ket{-,-} \right),
            \label{eq:cluster_state}
            \end{equation}
            built from the paired branch configurations
            \begin{equation}
                \ket{\sigma,\tau} = \ket{\sigma\alpha,\sigma\alpha, \tau\alpha,\tau\alpha},
                \qquad \sigma,\tau=\pm .
            \label{eq:cluster_branches}
            \end{equation}}  
        
        The paired branch manifold is selected by
            \begin{equation}
                \begin{split}
                    H_{\mathrm{pair}}^{(C)}={}& \lambda_{12} (a_1-a_2)^\dagger(a_1-a_2) \\
                    &+ \lambda_{34} (a_3-a_4)^\dagger(a_3-a_4),
                \end{split}
                \qquad \lambda_{12},\lambda_{34}>0 .
            \label{eq:cluster_pair}
            \end{equation}
        Within the branch manifold, its zero-energy condition imposes $s_1=s_2$ and $s_3=s_4$, reducing the original sixteen branch configurations to the four states in Eq.~\eqref{eq:cluster_branches}.
        
        To resolve the remaining degeneracy, we introduce branch-sign operators
            \begin{equation}
                S_A=\frac{a_1+a_2}{2\alpha}, 
                \qquad S_B=\frac{a_3+a_4}{2\alpha},
            \label{eq:cluster_sign}
            \end{equation}
        and pair-flip operators
            \begin{equation}
                F_A=\Pi_1\Pi_2, 
                \qquad F_B=\Pi_3\Pi_4.
            \label{eq:cluster_flip}
            \end{equation}
        On the selected branch states, $S_A$ and $S_B$ return the signs $\sigma$ and $\tau$, whereas $F_A$ and $F_B$ reverse the corresponding branch pair. The required graph correlations are imposed by
            \begin{equation}
                K_A=F_A S_B,
                \qquad K_B=S_A F_B,
            \label{eq:cluster_correlations}
            \end{equation}
        for which
            \begin{equation}
                K_A\ket{C}=K_B\ket{C}=\ket{C}.
            \end{equation}
        A positive-semidefinite correlation selector is therefore
            \begin{equation}
                \begin{split}
                    H_{\mathrm{corr}}^{(C)} ={}& \mu_A \left(\mathbb I-K_A\right)^\dagger \left(\mathbb I-K_A\right) \\
                    &+ \mu_B \left(\mathbb I-K_B\right)^\dagger \left(\mathbb I-K_B\right),
                \end{split}
            \qquad \mu_A,\mu_B>0 .
            \label{eq:cluster_selector}
            \end{equation}
        
        The complete cluster parent Hamiltonian is
            \begin{equation}
                H_C = H_{\mathrm{br}} + H_{\mathrm{pair}}^{(C)} + H_{\mathrm{corr}}^{(C)}.
            \label{eq:cluster_parent}
            \end{equation}
        The branch and pair constraints first select the four-dimensional manifold in Eq.~\eqref{eq:cluster_branches}, while the two graph constraints isolate $\ket{C}$ as their unique { common zero-energy state}, $16 \longrightarrow 4 \longrightarrow 1$. A complete uniqueness proof is given in the Supplemental Material~\cite{SM}.
        
        { The construction extends to larger graph-state resources whenever the corresponding bosonic constraints have a one-dimensional common kernel within the selected branch manifold.}

        \emph{W-type cat states.-}\label{sec:w_bosonic} The W-type construction illustrates a different selection mechanism. Rather than imposing a parity sector or graph-state correlations, it first fixes the number of positive coherent branches and then selects the permutation-symmetric superposition within that sector.
        
        For $M$ modes, define
            \begin{equation}
                \ket{W_M} = \mathcal N_{W,M} \sum_{j=1}^{M} \ket{-\alpha,\ldots,+\alpha_j,\ldots,-\alpha},
            \label{eq:w_state}
            \end{equation}
        where the positive branch occupies each mode once and $\mathcal N_{W,M}$ is a normalization factor.
        
        The sector containing exactly one positive branch is selected by the nullifier
            \begin{equation}
                L_W^{(M)} = \sum_{j=1}^{M}a_j+(M-2)\alpha
            \label{eq:w_nullifier}
            \end{equation}
        and the positive-semidefinite constraint
            \begin{equation}
                H_{\mathrm{def}}^{(W)} = \lambda_W \left(L_W^{(M)}\right)^\dagger L_W^{(M)},
                \qquad \lambda_W>0.
            \label{eq:w_defect}
            \end{equation}
        Indeed, for a branch configuration $\ket{\bm s}=\ket{s_1\alpha,\ldots,s_M\alpha}$,
            \begin{equation}
                L_W^{(M)}\ket{\bm s} = \alpha \left[ \sum_{j=1}^{M}s_j+(M-2) \right] \ket{\bm s}.
            \label{eq:w_nullifier_action}
            \end{equation}
        Its zero-energy condition therefore requires $\sum_j s_j=-(M-2)$ and retains precisely the $M$ configurations appearing in Eq.~\eqref{eq:w_state}.
        
        Their symmetric superposition is selected by mode-exchange symmetry. Let $\mathsf S_{ij}$ denote the bosonic swap operator, $\mathsf S_{ij} \ket{\ldots,\beta_i,\ldots,\beta_j,\ldots} = \ket{\ldots,\beta_j,\ldots,\beta_i,\ldots}$, with $\mathsf S_{ij}^\dagger=\mathsf S_{ij}$ and $\mathsf S_{ij}^2=\mathbb I$. For any connected set of mode pairs $\mathcal E_W$, define
            \begin{equation}
                H_{\mathrm{sym}}^{(W)} = \kappa \sum_{(i,j)\in\mathcal E_W} \left( \mathbb I-\mathsf S_{ij} \right),
                \qquad \kappa>0.
            \label{eq:w_symmetry}
            \end{equation}
        Each term in Eq.~\eqref{eq:w_symmetry} is positive semidefinite, because $\mathsf S_{ij}$ is Hermitian and has eigenvalues $\pm1$. Although the swap is not a polynomial in the local oscillator operators, its expectation value is experimentally accessible. Introducing the symmetric and antisymmetric modes
            \begin{equation}
                b_{ij,\pm} = \frac{a_i\pm a_j}{\sqrt{2}},
            \end{equation}
        the exchange $i\leftrightarrow j$ leaves $b_{ij,+}$ unchanged and transforms $b_{ij,-}\rightarrow-b_{ij,-}$. Consequently,
            \begin{equation}
                \mathsf S_{ij} = (-1)^{b_{ij,-}^{\dagger}b_{ij,-}}.
            \label{eq:swap_parity}
            \end{equation}
        Thus, in platforms supporting mode mixing and parity detection, $\langle\mathsf S_{ij}\rangle$ can be obtained by interfering modes $i$ and $j$ and measuring the parity of the antisymmetric output mode. Here $\mathsf S_{ij}$ is used as an exact symmetry selector; the physical realization of $\mathbb I-\mathsf S_{ij}$ as an interaction remains platform-dependent.
        
        Within the single-positive-branch sector, the zero-energy condition $\mathsf S_{ij}\ket{\psi}=\ket{\psi}$ requires equal amplitudes for the defect positions connected by the edge $(i,j)$. A connected set $\mathcal E_W$ therefore fixes all amplitudes to be equal and selects $\ket{W_M}$ uniquely.
        
        The complete W-type parent Hamiltonian is
            \begin{equation}
                H_W = H_{\mathrm{br}} + H_{\mathrm{def}}^{(W)} + H_{\mathrm{sym}}^{(W)}.
            \label{eq:w_parent}
            \end{equation}
        The successive constraints reduce the zero-energy manifold according to $2^M \longrightarrow M \longrightarrow 1$. Thus the W-type state is selected by a fixed-branch-number constraint followed by permutation symmetrization, rather than by a parity or graph-stabilizer sector. The uniqueness proof and the derivation of the swap-parity measurement relation are given in the Supplemental Material~\cite{SM}.

    \paragraph{Encoded-qubit representation---}\label{sec:encoded_qubits}For well-separated coherent branches, $\braket{\alpha|-\alpha} = e^{-2|\alpha|^2}  \ll 1$, the two states $\ket{\pm\alpha}$ define an approximate logical-qubit basis,
        \begin{equation}
            \ket{0_L}=\ket{-\alpha}, 
            \qquad  \ket{1_L}=\ket{\alpha}.
        \label{eq:logical_basis}
        \end{equation}
    We denote the corresponding Pauli operators by $X_j$, $Y_j$, and $Z_j$, with
        \begin{equation}
            Z_j\ket{\pm\alpha}_j = \pm\ket{\pm\alpha}_j, 
            \qquad X_j\ket{\pm\alpha}_j = \ket{\mp\alpha}_j.
        \end{equation}
    Within the branch manifold, the bosonic operators entering the parent Hamiltonians then reduce to the logical actions summarized in Table~\ref{tab:operator_dictionary}. Corrections to this mapping are controlled by the finite branch overlap $e^{-2|\alpha|^2}$.

        \begin{table}[t]
    \caption{
    Encoded-qubit actions of bosonic operators within the
    coherent-state branch manifold.
    }
    \label{tab:operator_dictionary}
    \centering

    \setlength{\tabcolsep}{3pt}
    \renewcommand{\arraystretch}{1.15}

    \resizebox{\columnwidth}{!}{%
    \begin{tabular}{|c|c|c|c|c|c|}
        \hline
        \multicolumn{6}{|c|}{\textbf{Bosonic operator}}
        \\ \hline

        $(-1)^{a_j^\dagger a_j}$ & $a_j/\alpha$ & $\bigotimes_j\Pi_j$ & $(a_i-a_j)^\dagger(a_i-a_j)$ & $L_W^{(M)}/\alpha$ & $\mathsf S_{ij}$\\ \hline \hline
        \multicolumn{6}{|c|}{\textbf{Encoded action}}
        \\ \hline
        $X_j$ & $Z_j$ & $\bigotimes_jX_j$ & $2|\alpha|^2(\mathbb I-Z_iZ_j)$ & $\sum_j Z_j+(M-2)\mathbb I$ & $\operatorname{SWAP}_{ij}$
        \\ \hline

    \end{tabular}%
    }
\end{table}

    The logical swap appearing in the W construction can be expressed as
        \begin{equation}
            \operatorname{SWAP}_{ij} = \frac{1}{2} \left( \mathbb I + X_iX_j + Y_iY_j + Z_iZ_j \right).
        \label{eq:logical_swap}
        \end{equation}
    The three bosonic constructions therefore reduce to familiar correlation-, stabilizer-, and exchange-based parent Hamiltonians.

    For the GHZ-type cat, the coherent branches become $\ket{\alpha}^{\otimes M}\leftrightarrow\ket{1}^{\otimes M}$ and $\ket{-\alpha}^{\otimes M}\leftrightarrow\ket{0}^{\otimes M}$. Restricting Eq.~\eqref{eq:ghz_parent} to the branch manifold gives
        \begin{equation}
            \begin{split}
                H_{\mathrm{GHZ},\pm}^{\mathrm{eff}} ={}& 2|\alpha|^2 \sum_{(i,j)\in\mathcal E} \lambda_{ij} \left( \mathbb I-Z_iZ_j \right) \\
                &+ \gamma \left( \mathbb I \mp \bigotimes_{j=1}^{M}X_j \right).
            \end{split}
        \label{eq:ghz_effective}
        \end{equation}
    The first term enforces equal logical branch signs across the connected mode graph, while the second selects the global $X$-symmetry sector.

    For the cluster construction, the pair-aligned sectors define two encoded qubits, $A=(1,2)$ and $B=(3,4)$. Inside this reduced manifold, we identify $\overline Z_A=Z_1=Z_2$, $\overline X_A=X_1X_2$, and analogously $       \overline Z_B=Z_3=Z_4$, $\overline X_B=X_3X_4$. The effective cluster Hamiltonian is then
        \begin{equation}
            \begin{split}
                H_C^{\mathrm{eff}} ={}& 2|\alpha|^2 \bigl[ \lambda_{12}(\mathbb I-Z_1Z_2) + \lambda_{34}(\mathbb I-Z_3Z_4) \bigr] \\
                &+ \mu_A \left( \mathbb I-\overline X_A\overline Z_B \right)^2 + \mu_B \left(\mathbb I-\overline Z_A\overline X_B \right)^2 .
            \end{split}
        \label{eq:cluster_effective}
        \end{equation}
    After the pair-alignment constraints are satisfied, the remaining terms are the two graph-state stabilizer penalties. Since the stabilizers square to the identity, these squared terms are equivalent, up to an overall positive factor, to the standard cluster-state parent Hamiltonian.

    Finally, the W branch sector maps to the one-excitation logical subspace,
        \begin{equation}
            \ket{W_M} \longrightarrow \frac{1}{\sqrt M} \sum_{j=1}^{M} \ket{0\cdots 1_j\cdots0}.
        \label{eq:w_logical_state}
        \end{equation}
    Using Table~\ref{tab:operator_dictionary}, the corresponding effective Hamiltonian is
        \begin{equation}
            \begin{split}
                H_W^{\mathrm{eff}} ={}& \lambda_W|\alpha|^2 \left[ \sum_{j=1}^{M}Z_j + (M-2)\mathbb I \right]^2 \\
                &+ \kappa \sum_{(i,j)\in\mathcal E_W} \left( \mathbb I-\operatorname{SWAP}_{ij}\right).
            \end{split}
        \label{eq:w_effective}
        \end{equation}
    The first term selects the one-excitation sector, while the swap penalties select its permutation-symmetric component. For connected $\mathcal E_W$, their unique common zero-energy state is the logical W state.

    \paragraph{Generation, stabilization, and certification---}\label{sec:generation_stabilization_certification}
    
    The common-kernel structure of the parent Hamiltonians also provides a basis for coherent preparation, dissipative stabilization, and state certification. These tasks need not employ the same physical interactions, but they are governed by the same defining constraints.

    For coherent preparation, an effective Hamiltonian reproducing the low-energy structure of $H_{\mathrm{target}}$ may be combined with cooling or adiabatic deformation toward its ground state. The relative weights of the constraints determine the excitation spectrum and hence the energetic protection of the target state.
    
    A complementary dissipative construction follows whenever a constraint can be written as $H_\mu=A_\mu^\dagger A_\mu$, with $A_\mu$ used as an engineered Lindblad jump operator,
    \begin{align*}
        \dot\rho&=-i[H,\rho] +\sum_\mu\kappa_\mu\mathcal D[A_\mu]\rho,\\
         \mathcal D[A]\rho &=A\rho A^\dagger -\frac{1}{2}\{A^\dagger A,\rho\}.
    \end{align*}
    A state satisfying $A_\mu\ket{\psi_0}=0$ for every $\mu$ is a common dark state of the dissipative channels. For it to be stationary under the full dynamics, the Hamiltonian must also preserve it, for example, $H\ket{\psi_0}=E_0\ket{\psi_0}$.
    
    For the branch constraint, $A_{\mathrm{br},j}=a_j^2-\alpha^2$, two-photon dissipation confines each oscillator to $\operatorname{span}\{\ket{\alpha},\ket{-\alpha}\}$. The remaining constraints then distinguish the required multimode stabilization: branch alignment and parity selection for GHZ-type states, graph correlations for cluster states, and fixed-branch-number selection with symmetrization for W-type states.
        
    The parent Hamiltonian also provides a natural certification observable. If $\ket{\psi_0}$ is its unique zero-energy state and $\Delta$ is the spectral gap above it, then 
        \begin{equation}
            H_{\mathrm{target}} \geq \Delta \left( \mathbb I - \ket{\psi_0}\!\bra{\psi_0} \right),
        \label{eq:fidelity_operator_bound}
        \end{equation}
    which yields the fidelity bound
        \begin{equation}
            \bra{\psi_0}\rho\ket{\psi_0} \geq 1- \frac{ \Tr\!\left( \rho H_{\mathrm{target}} \right) }{\Delta}.
        \label{eq:fidelity_energy_bound}
        \end{equation}
    Thus a small parent-Hamiltonian energy certifies proximity to the target state without requiring full state tomography.

    For polynomial constraints, the parent energy can be reconstructed from suitable field moments and interpreted in terms of generalized nonlinear variances. The swap contribution in the W construction is instead accessible through mode mixing and parity detection. More specific properties can be certified by comparing the measured energy with bounds optimized over restricted state classes, such as separable, Gaussian, or bounded-complexity states. Representative expectation-value comparisons for the target and orthogonal symmetry sectors are given in the Supplemental Material~\cite{SM}.

    \paragraph{Conclusion---}\label{sec:conclusion}
    We have introduced a constraint-based construction of parent Hamiltonians for entangled multimode bosonic cat resources. Local branch constraints establish the coherent-state manifold, while correlation and symmetry constraints select a unique global state within it. GHZ-, cluster-, and W-type cats demonstrate alignment-, graph-, and permutation-based selection mechanisms, and for well-separated branches the resulting oscillator Hamiltonians reduce to encoded-qubit correlation, stabilizer, and exchange models. The framework extends to broader graph-state resources whenever a complete set of bosonic constraints yields a unique common kernel. Beyond state definition, the same kernel provides a unified basis for coherent preparation, dissipative stabilization, and energy- or moment-based certification. This provides a systematic route from bosonic resource states to the constraints required for their physical realization and verification.

    \paragraph{Acknowledgments---}The authors acknowledge fruitful discussions with Petr Marek, Michal Matul\'ik, and Ananga Mohan Datta. This work was supported by the Carlsberg Foundation through the Semper Ardens project QCooL.

\clearpage
\onecolumngrid

\begin{center}
    {\large\bfseries
    Supplemental Material for: Symmetry-resolved parent Hamiltonians for entangled bosonic cat resources
    }

    \vspace{0.5em}

    Šimon Bräuer and Klaus Mølmer

    \vspace{0.3em}

    Niels Bohr Institute, University of Copenhagen,
    Jagtvej 155 A, DK-2200 Copenhagen, Denmark
\end{center}

\vspace{1em}

\setcounter{equation}{0}
\renewcommand{\theequation}{S\arabic{equation}}

\setcounter{figure}{0}
\renewcommand{\thefigure}{S\arabic{figure}}

\setcounter{table}{0}
\renewcommand{\thetable}{S\arabic{table}}

\section{Constraint completeness and uniqueness}
\label{sec:supplement_constraint_completeness}

In this section, we establish the common-kernel criterion underlying
the parent-Hamiltonian construction and prove that the GHZ-, cluster-,
and W-type cat states discussed in the main text are the unique
zero-energy states of their respective Hamiltonians.

Throughout the Supplemental Material, we assume
$\alpha\neq0$ and strictly positive coefficients multiplying all
constraint terms. The nonzero-amplitude condition ensures that the
coherent branches $\ket{\alpha}$ and $\ket{-\alpha}$ are distinct.

\subsection{General common-kernel criterion}
\label{sec:supplement_general_kernel}

Consider a Hamiltonian of the form
\begin{equation}
    \hat H
    =
    \sum_{\mu=0}^{N}
    \lambda_{\mu}\hat H_{\mu},
    \qquad
    \hat H_{\mu}\geq0,
    \qquad
    \lambda_{\mu}>0.
\label{eq:supplement_general_hamiltonian}
\end{equation}
For the parent-Hamiltonian constructions considered below, the positive-semidefinite terms are chosen to possess a nontrivial common kernel containing the desired target state. Its kernel is the intersection of the kernels of the individual
positive-semidefinite terms,
\begin{equation}
    \ker\hat H
    =
    \bigcap_{\mu=0}^{N}
    \ker\hat H_{\mu}.
\label{eq:supplement_kernel_intersection}
\end{equation}
Every vector belonging to all individual kernels clearly belongs to
$\ker\hat H$. Conversely, let
$\ket{\psi}\in\ker\hat H$. Then
\begin{equation}
    0
    =
    \bra{\psi}\hat H\ket{\psi}
    =
    \sum_{\mu=0}^{N}
    \lambda_{\mu}
    \bra{\psi}\hat H_{\mu}\ket{\psi}.
\label{eq:supplement_zero_expectation}
\end{equation}
Every term in this sum is nonnegative. Their sum can therefore vanish
only when
\begin{equation}
    \bra{\psi}\hat H_{\mu}\ket{\psi}=0
    \qquad
    \text{for every }\mu.
\label{eq:supplement_individual_zero_expectation}
\end{equation}
Since $\hat H_{\mu}\geq0$, this is equivalent to
\begin{equation}
    \hat H_{\mu}^{1/2}\ket{\psi}=0,
\end{equation}
and hence
\begin{equation}
    \hat H_{\mu}\ket{\psi}=0
    \qquad
    \text{for every }\mu.
\end{equation}
This proves Eq.~\eqref{eq:supplement_kernel_intersection}.

The parent Hamiltonians introduced in the main text have the structure
\begin{equation}
    \hat H_{\mathrm{target}}
    =
    \hat H_{\mathrm{br}}
    +
    \sum_{\mu}
    \lambda_{\mu}\hat H_{\mu},
\label{eq:supplement_target_hamiltonian}
\end{equation}
where the universal branch Hamiltonian is
\begin{equation}
    \hat H_{\mathrm{br}}
    =
    \sum_{j=1}^{M}
    \left(
        \hat a_j^2-\alpha^2
    \right)^{\dagger}
    \left(
        \hat a_j^2-\alpha^2
    \right).
\label{eq:supplement_branch_hamiltonian}
\end{equation}

We first establish the kernel of the corresponding single-mode
constraint. Let
\begin{equation}
    \ket{\psi}
    =
    \sum_{n=0}^{\infty}
    c_n\ket{n}
\end{equation}
satisfy
\begin{equation}
    \left(
        \hat a^2-\alpha^2
    \right)\ket{\psi}=0.
\label{eq:supplement_single_mode_nullifier}
\end{equation}
Expansion in the Fock basis gives
\begin{equation}
    \sum_{n=0}^{\infty}
    c_{n+2}
    \sqrt{(n+1)(n+2)}
    \ket{n}
    =
    \alpha^2
    \sum_{n=0}^{\infty}
    c_n\ket{n},
\end{equation}
and therefore the recurrence relation
\begin{equation}
    c_{n+2}
    =
    \frac{\alpha^2}{
        \sqrt{(n+1)(n+2)}
    }
    c_n.
\label{eq:supplement_recurrence}
\end{equation}
The even and odd coefficients are fixed independently:
\begin{align}
    c_{2m}
    &=
    c_0
    \frac{\alpha^{2m}}{\sqrt{(2m)!}},
\label{eq:supplement_even_coefficients}
    \\
    c_{2m+1}
    &=
    c_1
    \frac{\alpha^{2m}}{\sqrt{(2m+1)!}}.
\label{eq:supplement_odd_coefficients}
\end{align}
These two independent normalizable solutions are proportional to the
even and odd coherent-state superpositions
\begin{equation}
    \ket{\alpha}\pm\ket{-\alpha}.
\end{equation}
Equivalently,
\begin{equation}
    \ker
    \left[
        \left(
            \hat a^2-\alpha^2
        \right)^{\dagger}
        \left(
            \hat a^2-\alpha^2
        \right)
    \right]
    =
    \operatorname{span}
    \left\{
        \ket{\alpha},
        \ket{-\alpha}
    \right\}.
\label{eq:supplement_single_mode_kernel}
\end{equation}

Because the terms in
Eq.~\eqref{eq:supplement_branch_hamiltonian}
act locally on different modes, the multimode kernel is the tensor
product of the corresponding single-mode kernels,
\begin{equation}
    \ker\hat H_{\mathrm{br}}
    =
    \mathcal B_{\alpha}^{(M)}
    =
    \operatorname{span}
    \left\{
        \ket{\bm s}
        =
        \ket{s_1\alpha,\ldots,s_M\alpha}
        :
        s_j=\pm1
    \right\}.
\label{eq:supplement_multimode_branch_kernel}
\end{equation}
For $\alpha\neq0$, the two coherent states
$\ket{\alpha}$ and $\ket{-\alpha}$ are linearly independent.
Consequently, their $M$-mode tensor products are also linearly
independent, and
\begin{equation}
    \dim\mathcal B_{\alpha}^{(M)}
    =
    2^M.
\label{eq:supplement_branch_dimension}
\end{equation}

The remaining terms in
Eq.~\eqref{eq:supplement_target_hamiltonian}
act as constraints inside this branch manifold. If their common kernel
inside $\mathcal B_{\alpha}^{(M)}$ is one-dimensional, then
Eq.~\eqref{eq:supplement_kernel_intersection} guarantees that the full
Hamiltonian has a unique zero-energy ground state.

\subsubsection{Conditional extension to bosonic graph states}
\label{sec:supplement_graph_extension}

The same criterion gives the graph-state extension described in the
main text. Suppose that a set of bosonic operators
$\{\hat{\mathcal K}_{\mu}\}$ preserves a selected branch manifold
\begin{equation}
    \mathcal B_{\mathrm{sel}}
    \subseteq
    \mathcal B_{\alpha}^{(M)}
\end{equation}
and reproduces a complete set of graph-state stabilizers within that
manifold. Define the positive-semidefinite constraints
\begin{equation}
    \hat H_{\mu}^{(G)}
    =
    \left(
        \mathbb I-\hat{\mathcal K}_{\mu}
    \right)^{\dagger}
    \left(
        \mathbb I-\hat{\mathcal K}_{\mu}
    \right).
\label{eq:supplement_graph_penalty}
\end{equation}
Their common zero-energy condition is
\begin{equation}
    \hat{\mathcal K}_{\mu}\ket{\psi}
    =
    \ket{\psi}
    \qquad
    \text{for every }\mu.
\label{eq:supplement_graph_constraint}
\end{equation}

Let $\hat H_{\mathrm{sel}}\geq0$ denote any additional constraint whose
kernel inside $\mathcal B_{\alpha}^{(M)}$ is
$\mathcal B_{\mathrm{sel}}$. The corresponding graph-state parent
Hamiltonian is
\begin{equation}
    \hat H_G
    =
    \hat H_{\mathrm{br}}
    +
    \hat H_{\mathrm{sel}}
    +
    \sum_{\mu}
    \lambda_{\mu}
    \hat H_{\mu}^{(G)},
    \qquad
    \lambda_{\mu}>0.
\label{eq:supplement_graph_hamiltonian}
\end{equation}
If the simultaneous unit-eigenspace of the operators
$\hat{\mathcal K}_{\mu}$ inside
$\mathcal B_{\mathrm{sel}}$ is one-dimensional, then
\begin{equation}
    \ker\hat H_G
    =
    \operatorname{span}
    \left\{
        \ket{G_{\alpha}}
    \right\},
\label{eq:supplement_graph_unique_kernel}
\end{equation}
where $\ket{G_{\alpha}}$ is the corresponding bosonic graph state.

This is a conditional completeness statement: uniqueness follows once
the lifted bosonic constraints form a complete set inside the selected
coherent-branch manifold.

\subsection{GHZ-type cat state}
\label{sec:supplement_ghz}

The $M$-mode GHZ-type cat states are
\begin{equation}
    \ket{\mathrm{GHZ}_{\pm}}
    =
    \mathcal N_{\mathrm{GHZ},\pm}
    \left(
        \ket{\alpha}^{\otimes M}
        \pm
        \ket{-\alpha}^{\otimes M}
    \right),
\label{eq:supplement_ghz_state}
\end{equation}
with normalization
\begin{equation}
    \mathcal N_{\mathrm{GHZ},\pm}^{-2}
    =
    2
    \left(
        1
        \pm
        e^{-2M|\alpha|^2}
    \right).
\label{eq:supplement_ghz_normalization}
\end{equation}

The corresponding parent Hamiltonian is
\begin{equation}
    \hat H_{\mathrm{GHZ},\pm}
    =
    \hat H_{\mathrm{br}}
    +
    \hat H_{\mathrm{align}}^{(\mathrm{GHZ})}
    +
    \hat H_{\mathrm{par}}^{(\mathrm{GHZ},\pm)},
\label{eq:supplement_ghz_hamiltonian}
\end{equation}
where
\begin{equation}
    \hat H_{\mathrm{align}}^{(\mathrm{GHZ})}
    =
    \sum_{(i,j)\in\mathcal E}
    \lambda_{ij}
    \left(
        \hat a_i-\hat a_j
    \right)^{\dagger}
    \left(
        \hat a_i-\hat a_j
    \right),
    \qquad
    \lambda_{ij}>0,
\label{eq:supplement_ghz_alignment}
\end{equation}
and
\begin{equation}
    \hat H_{\mathrm{par}}^{(\mathrm{GHZ},\pm)}
    =
    \gamma
    \left(
        \mathbb I\mp\hat\Pi_G
    \right),
    \qquad
    \gamma>0.
\label{eq:supplement_ghz_parity_hamiltonian}
\end{equation}
The global photon-number parity is
\begin{equation}
    \hat\Pi_G
    =
    (-1)^{
        \sum_{j=1}^{M}
        \hat a_j^{\dagger}\hat a_j
    }
    =
    \prod_{j=1}^{M}
    \hat\Pi_j,
\qquad
    \hat\Pi_j
    =
    (-1)^{\hat a_j^{\dagger}\hat a_j}.
\label{eq:supplement_global_parity}
\end{equation}
We assume that the graph with edge set $\mathcal E$ is connected.

\subsubsection{Zero-energy property}
\label{sec:supplement_ghz_zero_energy}

The branch term annihilates both components of
Eq.~\eqref{eq:supplement_ghz_state}, since
\begin{equation}
    \hat a_j^2\ket{\pm\alpha}_j
    =
    \alpha^2\ket{\pm\alpha}_j.
\label{eq:supplement_ghz_branch_action}
\end{equation}
The two globally aligned components are also annihilated by every
alignment nullifier:
\begin{equation}
    \left(
        \hat a_i-\hat a_j
    \right)
    \ket{\pm\alpha}^{\otimes M}
    =
    0.
\label{eq:supplement_ghz_alignment_zero}
\end{equation}

The single-mode parity operator transforms a coherent state according
to
\begin{equation}
    \hat\Pi_j\ket{\alpha}_j
    =
    \ket{-\alpha}_j.
\label{eq:supplement_single_parity_action}
\end{equation}
Consequently, global parity exchanges the two aligned branches,
\begin{align}
    \hat\Pi_G\ket{\alpha}^{\otimes M}
    &=
    \ket{-\alpha}^{\otimes M},
\label{eq:supplement_global_parity_plus}
    \\
    \hat\Pi_G\ket{-\alpha}^{\otimes M}
    &=
    \ket{\alpha}^{\otimes M},
\label{eq:supplement_global_parity_minus}
\end{align}
and therefore
\begin{equation}
    \hat\Pi_G\ket{\mathrm{GHZ}_{\pm}}
    =
    \pm
    \ket{\mathrm{GHZ}_{\pm}}.
\label{eq:supplement_ghz_parity_eigenvalue}
\end{equation}

All three contributions to
Eq.~\eqref{eq:supplement_ghz_hamiltonian}
consequently annihilate the target state:
\begin{equation}
    \hat H_{\mathrm{GHZ},\pm}
    \ket{\mathrm{GHZ}_{\pm}}
    =
    0.
\label{eq:supplement_ghz_zero_energy}
\end{equation}

\subsubsection{Uniqueness}
\label{sec:supplement_ghz_uniqueness}

Let $\ket{\psi}$ be an arbitrary zero-energy state of
$\hat H_{\mathrm{GHZ},\pm}$. By the common-kernel criterion, it must
belong to the kernel of every positive-semidefinite contribution.

The branch constraint first implies
\begin{equation}
    \ket{\psi}
    =
    \sum_{\bm s}
    c_{\bm s}\ket{\bm s},
    \qquad
    \ket{\bm s}
    =
    \ket{s_1\alpha,\ldots,s_M\alpha},
    \qquad
    s_j=\pm1.
\label{eq:supplement_general_branch_state}
\end{equation}
For every edge $(i,j)\in\mathcal E$,
\begin{equation}
    \left(
        \hat a_i-\hat a_j
    \right)
    \ket{\bm s}
    =
    \alpha
    \left(
        s_i-s_j
    \right)
    \ket{\bm s}.
\label{eq:supplement_alignment_action}
\end{equation}
The zero-energy condition associated with the alignment Hamiltonian
requires
\begin{equation}
    \left(
        \hat a_i-\hat a_j
    \right)
    \ket{\psi}
    =
    0
    \qquad
    \text{for every }(i,j)\in\mathcal E.
\label{eq:supplement_alignment_condition}
\end{equation}
Using Eq.~\eqref{eq:supplement_general_branch_state}, this becomes
\begin{equation}
    \sum_{\bm s}
    c_{\bm s}
    \alpha(s_i-s_j)
    \ket{\bm s}
    =
    0.
\end{equation}
Since the branch states are linearly independent for $\alpha\neq0$,
we must have
\begin{equation}
    c_{\bm s}=0
    \qquad
    \text{whenever }
    s_i\neq s_j
\label{eq:supplement_alignment_coefficient_condition}
\end{equation}
for every edge $(i,j)\in\mathcal E$.

Because $\mathcal E$ is connected, satisfying
Eq.~\eqref{eq:supplement_alignment_coefficient_condition} on every
edge requires
\begin{equation}
    s_1=s_2=\cdots=s_M.
\label{eq:supplement_global_alignment_condition}
\end{equation}
The alignment constraints therefore reduce the state to
\begin{equation}
    \ket{\psi}
    =
    c_{+}\ket{\alpha}^{\otimes M}
    +
    c_{-}\ket{-\alpha}^{\otimes M}.
\label{eq:supplement_aligned_state}
\end{equation}

The remaining zero-energy condition is imposed by the parity term:
\begin{equation}
    \hat\Pi_G\ket{\psi}
    =
    \pm\ket{\psi}.
\label{eq:supplement_parity_condition}
\end{equation}
Using Eqs.~\eqref{eq:supplement_global_parity_plus} and
\eqref{eq:supplement_global_parity_minus}, the left-hand side is
\begin{equation}
    \hat\Pi_G\ket{\psi}
    =
    c_{+}\ket{-\alpha}^{\otimes M}
    +
    c_{-}\ket{\alpha}^{\otimes M}.
\end{equation}
Comparison with the right-hand side of
Eq.~\eqref{eq:supplement_parity_condition} gives
\begin{equation}
    c_{-}
    =
    \pm c_{+}.
\label{eq:supplement_ghz_coefficient_relation}
\end{equation}
Thus every zero-energy state is proportional to
$\ket{\mathrm{GHZ}_{\pm}}$, and
\begin{equation}
    \ker\hat H_{\mathrm{GHZ},\pm}
    =
    \operatorname{span}
    \left\{
        \ket{\mathrm{GHZ}_{\pm}}
    \right\}.
\label{eq:supplement_ghz_unique_kernel}
\end{equation}
The constraint hierarchy therefore produces the exact reduction
\begin{equation}
    2^M
    \longrightarrow
    2
    \longrightarrow
    1.
\label{eq:supplement_ghz_reduction}
\end{equation}

\subsection{Cluster-type cat state}
\label{sec:supplement_cluster}

We consider the four-mode paired branch configurations
\begin{equation}
    \ket{\sigma,\tau}
    =
    \ket{
        \sigma\alpha,
        \sigma\alpha,
        \tau\alpha,
        \tau\alpha
    },
    \qquad
    \sigma,\tau=\pm.
\label{eq:supplement_cluster_branches}
\end{equation}
The target cluster-type cat state is
\begin{equation}
    \ket{C}
    =
    \mathcal N_C
    \left(
        \ket{+,+}
        +
        \ket{+,-}
        +
        \ket{-,+}
        -
        \ket{-,-}
    \right).
\label{eq:supplement_cluster_state}
\end{equation}
For the sign pattern in Eq.~\eqref{eq:supplement_cluster_state},
all finite-overlap cross terms cancel, and therefore
\begin{equation}
    \mathcal N_C=\frac{1}{2}.
\label{eq:supplement_cluster_normalization}
\end{equation}

The corresponding parent Hamiltonian is
\begin{equation}
    \hat H_C
    =
    \hat H_{\mathrm{br}}
    +
    \hat H_{\mathrm{pair}}^{(C)}
    +
    \hat H_{\mathrm{corr}}^{(C)}.
\label{eq:supplement_cluster_hamiltonian}
\end{equation}
The pair-selection Hamiltonian is
\begin{align}
    \hat H_{\mathrm{pair}}^{(C)}
    ={}&
    \lambda_{12}
    \left(
        \hat a_1-\hat a_2
    \right)^{\dagger}
    \left(
        \hat a_1-\hat a_2
    \right)
    \nonumber\\
    &+
    \lambda_{34}
    \left(
        \hat a_3-\hat a_4
    \right)^{\dagger}
    \left(
        \hat a_3-\hat a_4
    \right),
\label{eq:supplement_cluster_pair_hamiltonian}
\end{align}
where
\begin{equation}
    \lambda_{12},\lambda_{34}>0.
\end{equation}

Inside the pair-selected branch manifold, we introduce the branch-sign
operators
\begin{equation}
    \hat S_A
    =
    \frac{
        \hat a_1+\hat a_2
    }{2\alpha},
    \qquad
    \hat S_B
    =
    \frac{
        \hat a_3+\hat a_4
    }{2\alpha},
\label{eq:supplement_cluster_sign_operators}
\end{equation}
and the pair-flip operators
\begin{equation}
    \hat F_A
    =
    \hat\Pi_1\hat\Pi_2,
    \qquad
    \hat F_B
    =
    \hat\Pi_3\hat\Pi_4.
\label{eq:supplement_cluster_flip_operators}
\end{equation}
The graph-correlation operators are
\begin{equation}
    \hat K_A
    =
    \hat F_A\hat S_B,
    \qquad
    \hat K_B
    =
    \hat S_A\hat F_B.
\label{eq:supplement_cluster_correlation_operators}
\end{equation}
The associated positive-semidefinite correlation Hamiltonian is
\begin{align}
    \hat H_{\mathrm{corr}}^{(C)}
    ={}&
    \mu_A
    \left(
        \mathbb I-\hat K_A
    \right)^{\dagger}
    \left(
        \mathbb I-\hat K_A
    \right)
    \nonumber\\
    &+
    \mu_B
    \left(
        \mathbb I-\hat K_B
    \right)^{\dagger}
    \left(
        \mathbb I-\hat K_B
    \right),
\label{eq:supplement_cluster_correlation_hamiltonian}
\end{align}
with
\begin{equation}
    \mu_A,\mu_B>0.
\end{equation}

\subsubsection{Pair-selected branch manifold}
\label{sec:supplement_cluster_pair_manifold}

A general state in the kernel of the four-mode branch Hamiltonian can
be written as
\begin{equation}
    \ket{\psi}
    =
    \sum_{s_1,s_2,s_3,s_4=\pm1}
    c_{s_1s_2s_3s_4}
    \ket{
        s_1\alpha,
        s_2\alpha,
        s_3\alpha,
        s_4\alpha
    }.
\label{eq:supplement_cluster_general_branch_state}
\end{equation}
On an individual branch configuration,
\begin{align}
    \left(
        \hat a_1-\hat a_2
    \right)
    \ket{s_1\alpha,s_2\alpha,s_3\alpha,s_4\alpha}
    &=
    \alpha(s_1-s_2)
    \ket{s_1\alpha,s_2\alpha,s_3\alpha,s_4\alpha},
    \\
    \left(
        \hat a_3-\hat a_4
    \right)
    \ket{s_1\alpha,s_2\alpha,s_3\alpha,s_4\alpha}
    &=
    \alpha(s_3-s_4)
    \ket{s_1\alpha,s_2\alpha,s_3\alpha,s_4\alpha}.
\end{align}
Because the branch configurations are linearly independent, the
zero-energy conditions associated with
Eq.~\eqref{eq:supplement_cluster_pair_hamiltonian} require
\begin{equation}
    s_1=s_2,
    \qquad
    s_3=s_4.
\label{eq:supplement_cluster_pair_conditions}
\end{equation}
The sixteen-dimensional branch manifold is therefore reduced to
\begin{equation}
    \mathcal B_C
    =
    \operatorname{span}
    \left\{
        \ket{\sigma,\tau}:
        \sigma,\tau=\pm1
    \right\},
\label{eq:supplement_cluster_selected_manifold}
\end{equation}
with
\begin{equation}
    \dim\mathcal B_C=4.
\end{equation}

\subsubsection{Action of the graph-correlation operators}
\label{sec:supplement_cluster_operator_action}

The branch-sign operators act diagonally on the selected configurations:
\begin{equation}
    \hat S_A\ket{\sigma,\tau}
    =
    \sigma\ket{\sigma,\tau},
    \qquad
    \hat S_B\ket{\sigma,\tau}
    =
    \tau\ket{\sigma,\tau}.
\label{eq:supplement_cluster_sign_action}
\end{equation}
The pair parities reverse the signs of their corresponding coherent
branches,
\begin{equation}
    \hat F_A\ket{\sigma,\tau}
    =
    \ket{-\sigma,\tau},
    \qquad
    \hat F_B\ket{\sigma,\tau}
    =
    \ket{\sigma,-\tau}.
\label{eq:supplement_cluster_flip_action}
\end{equation}
Consequently,
\begin{equation}
    \hat K_A\ket{\sigma,\tau}
    =
    \tau\ket{-\sigma,\tau},
    \qquad
    \hat K_B\ket{\sigma,\tau}
    =
    \sigma\ket{\sigma,-\tau}.
\label{eq:supplement_cluster_correlation_action}
\end{equation}

\subsubsection{Zero-energy property}
\label{sec:supplement_cluster_zero_energy}

Every branch appearing in Eq.~\eqref{eq:supplement_cluster_state}
belongs to the kernel of both
$\hat H_{\mathrm{br}}$ and
$\hat H_{\mathrm{pair}}^{(C)}$.

Using Eq.~\eqref{eq:supplement_cluster_correlation_action}, one finds
\begin{equation}
    \hat K_A\ket{C}
    =
    \mathcal N_C
    \left(
    \ket{-,+}
    -\ket{-,-}
    +\ket{+,+}
    +\ket{+,-}
    \right)
    =
    \ket{C}.
\end{equation}
where the first line is understood after collecting the transformed
branch configurations in the ordering used in
Eq.~\eqref{eq:supplement_cluster_state}. Similarly,
\begin{equation}
    \hat K_B\ket{C}
    =
    \ket{C}.
\end{equation}
Equivalently, the target satisfies
\begin{equation}
    \left(
        \mathbb I-\hat K_A
    \right)\ket{C}
    =
    0,
    \qquad
    \left(
        \mathbb I-\hat K_B
    \right)\ket{C}
    =
    0.
\label{eq:supplement_cluster_nullifier_conditions}
\end{equation}
All contributions to
Eq.~\eqref{eq:supplement_cluster_hamiltonian} therefore annihilate the
target state,
\begin{equation}
    \hat H_C\ket{C}=0.
\label{eq:supplement_cluster_zero_energy}
\end{equation}

\subsubsection{Uniqueness}
\label{sec:supplement_cluster_uniqueness}

Let an arbitrary state in the pair-selected manifold be
\begin{equation}
    \ket{\psi}
    =
    \sum_{\sigma,\tau=\pm1}
    c_{\sigma,\tau}
    \ket{\sigma,\tau}.
\label{eq:supplement_cluster_general_selected_state}
\end{equation}
The zero-energy condition associated with the first graph constraint is
\begin{equation}
    \hat K_A\ket{\psi}
    =
    \ket{\psi}.
\label{eq:supplement_cluster_KA_condition}
\end{equation}
Using Eq.~\eqref{eq:supplement_cluster_correlation_action}, comparison
of the coefficients of each linearly independent branch state gives
\begin{equation}
    c_{\sigma,\tau}
    =
    \tau\,c_{-\sigma,\tau}.
\label{eq:supplement_cluster_KA_coefficients}
\end{equation}
Similarly,
\begin{equation}
    \hat K_B\ket{\psi}
    =
    \ket{\psi}
\label{eq:supplement_cluster_KB_condition}
\end{equation}
implies
\begin{equation}
    c_{\sigma,\tau}
    =
    \sigma\,c_{\sigma,-\tau}.
\label{eq:supplement_cluster_KB_coefficients}
\end{equation}

Choosing
\begin{equation}
    c_{+,+}=c
\end{equation}
as the free coefficient,
Eq.~\eqref{eq:supplement_cluster_KA_coefficients} gives
\begin{equation}
    c_{-,+}=c,
\end{equation}
while Eq.~\eqref{eq:supplement_cluster_KB_coefficients} gives
\begin{equation}
    c_{+,-}=c.
\end{equation}
Applying either constraint to the remaining coefficient yields
\begin{equation}
    c_{-,-}=-c.
\end{equation}
Thus the only simultaneous solution is
\begin{equation}
    \ket{\psi}
    =
    c
    \left(
        \ket{+,+}
        +
        \ket{+,-}
        +
        \ket{-,+}
        -
        \ket{-,-}
    \right),
\end{equation}
which is proportional to the target state
Eq.~\eqref{eq:supplement_cluster_state}. Therefore,
\begin{equation}
    \ker\hat H_C
    =
    \operatorname{span}
    \left\{
        \ket{C}
    \right\}.
\label{eq:supplement_cluster_unique_kernel}
\end{equation}
The complete constraint hierarchy produces the exact reduction
\begin{equation}
    16
    \longrightarrow
    4
    \longrightarrow
    1.
\label{eq:supplement_cluster_reduction}
\end{equation}

\subsection{W-type cat state}
\label{sec:supplement_w}

For $M$ modes, define the branch configurations
\begin{equation}
    \ket{j}
    =
    \ket{
        -\alpha,\ldots,
        +\alpha_j,
        \ldots,-\alpha
    },
    \qquad
    j=1,\ldots,M,
\label{eq:supplement_w_branches}
\end{equation}
where the positive coherent branch occupies mode $j$. The
permutation-symmetric W-type cat state is
\begin{equation}
    \ket{W_M}
    =
    \mathcal N_{W,M}
    \sum_{j=1}^{M}\ket{j}.
\label{eq:supplement_w_state}
\end{equation}

Two distinct configurations $\ket{i}$ and $\ket{j}$ differ in the
signs of two coherent amplitudes. Consequently,
\begin{equation}
    \braket{i|j}
    =
    e^{-4|\alpha|^2},
    \qquad
    i\neq j,
\label{eq:supplement_w_branch_overlap}
\end{equation}
and the normalization factor is
\begin{equation}
    \mathcal N_{W,M}^{-2}
    =
    M
    \left[
        1+(M-1)e^{-4|\alpha|^2}
    \right].
\label{eq:supplement_w_normalization}
\end{equation}

The complete W-type parent Hamiltonian is
\begin{equation}
    \hat H_W
    =
    \hat H_{\mathrm{br}}
    +
    \hat H_{\mathrm{def}}^{(W)}
    +
    \hat H_{\mathrm{sym}}^{(W)}.
\label{eq:supplement_w_hamiltonian}
\end{equation}
The fixed-defect constraint is
\begin{equation}
    \hat H_{\mathrm{def}}^{(W)}
    =
    \lambda_W
    \left(
        \hat L_W^{(M)}
    \right)^{\dagger}
    \hat L_W^{(M)},
    \qquad
    \lambda_W>0,
\label{eq:supplement_w_defect_hamiltonian}
\end{equation}
where
\begin{equation}
    \hat L_W^{(M)}
    =
    \sum_{j=1}^{M}\hat a_j
    +
    (M-2)\alpha.
\label{eq:supplement_w_nullifier}
\end{equation}

The permutation-symmetry constraint is
\begin{equation}
    \hat H_{\mathrm{sym}}^{(W)}
    =
    \kappa
    \sum_{(i,j)\in\mathcal E_W}
    \left(
        \mathbb I-\hat{\mathsf S}_{ij}
    \right),
    \qquad
    \kappa>0,
\label{eq:supplement_w_symmetry_hamiltonian}
\end{equation}
where $\hat{\mathsf S}_{ij}$ exchanges oscillator modes $i$ and $j$,
\begin{equation}
    \hat{\mathsf S}_{ij}
    \ket{\ldots,\beta_i,\ldots,\beta_j,\ldots}
    =
    \ket{\ldots,\beta_j,\ldots,\beta_i,\ldots}.
\label{eq:supplement_swap_definition}
\end{equation}
We assume that the graph with edge set $\mathcal E_W$ is connected.

\subsubsection{Fixed-defect branch manifold}
\label{sec:supplement_w_defect_manifold}

Consider an arbitrary coherent branch configuration
\begin{equation}
    \ket{\bm s}
    =
    \ket{s_1\alpha,\ldots,s_M\alpha},
    \qquad
    s_j=\pm1.
\end{equation}
The nullifier in Eq.~\eqref{eq:supplement_w_nullifier} acts diagonally:
\begin{equation}
    \hat L_W^{(M)}
    \ket{\bm s}
    =
    \alpha
    \left[
        \sum_{j=1}^{M}s_j+(M-2)
    \right]
    \ket{\bm s}.
\label{eq:supplement_w_nullifier_action}
\end{equation}
Its zero-energy condition is therefore
\begin{equation}
    \sum_{j=1}^{M}s_j
    =
    -(M-2).
\label{eq:supplement_w_sign_condition}
\end{equation}

Let $n_{+}$ denote the number of positive branch signs. Since
\begin{equation}
    \sum_{j=1}^{M}s_j
    =
    n_{+}-(M-n_{+})
    =
    2n_{+}-M,
\end{equation}
Eq.~\eqref{eq:supplement_w_sign_condition} implies
\begin{equation}
    2n_{+}-M
    =
    -(M-2),
\end{equation}
and hence
\begin{equation}
    n_{+}=1.
\label{eq:supplement_w_one_positive_branch}
\end{equation}
The combined branch and defect constraints therefore reduce the
$2^M$-dimensional branch manifold to
\begin{equation}
    \mathcal B_W
    =
    \operatorname{span}
    \left\{
        \ket{1},\ldots,\ket{M}
    \right\},
\label{eq:supplement_w_selected_manifold}
\end{equation}
with
\begin{equation}
    \dim\mathcal B_W=M.
\end{equation}

\subsubsection{Positivity of the swap constraint}
\label{sec:supplement_w_swap_positivity}

The mode-swap operator is Hermitian and unitary,
\begin{equation}
    \hat{\mathsf S}_{ij}^{\dagger}
    =
    \hat{\mathsf S}_{ij},
    \qquad
    \hat{\mathsf S}_{ij}^{2}
    =
    \mathbb I.
\label{eq:supplement_swap_properties}
\end{equation}
Its eigenvalues are therefore $\pm1$. It follows that
\begin{equation}
    \mathbb I-\hat{\mathsf S}_{ij}
    \geq0,
\label{eq:supplement_swap_penalty_positive}
\end{equation}
with eigenvalue zero on states symmetric under exchange of modes
$i$ and $j$, and eigenvalue two on antisymmetric states. Consequently,
$\hat H_{\mathrm{sym}}^{(W)}$ is positive semidefinite.

Equivalently, because
$\hat{\mathsf S}_{ij}^{\dagger}\hat{\mathsf S}_{ij}=\mathbb I$,
\begin{equation}
    \mathbb I-\hat{\mathsf S}_{ij}
    =
    \frac{1}{2}
    \left(
        \mathbb I-\hat{\mathsf S}_{ij}
    \right)^{\dagger}
    \left(
        \mathbb I-\hat{\mathsf S}_{ij}
    \right).
\label{eq:supplement_swap_nullifier_form}
\end{equation}
The swap term therefore also has the standard
positive-semidefinite nullifier form.

\subsubsection{Zero-energy property}
\label{sec:supplement_w_zero_energy}

Every branch $\ket{j}$ belongs to the kernel of
$\hat H_{\mathrm{br}}$ and satisfies the fixed-defect condition
\begin{equation}
    \hat L_W^{(M)}\ket{j}=0.
\label{eq:supplement_w_defect_zero}
\end{equation}
Therefore,
\begin{equation}
    \hat H_{\mathrm{br}}\ket{W_M}=0,
    \qquad
    \hat H_{\mathrm{def}}^{(W)}\ket{W_M}=0.
\end{equation}

A swap exchanges the positions of the positive branch,
\begin{equation}
    \hat{\mathsf S}_{ij}\ket{i}
    =
    \ket{j},
    \qquad
    \hat{\mathsf S}_{ij}\ket{j}
    =
    \ket{i},
\label{eq:supplement_w_swap_branch_action}
\end{equation}
while leaving all $\ket{k}$ with $k\neq i,j$ unchanged. Hence
\begin{equation}
    \hat{\mathsf S}_{ij}
    \sum_{k=1}^{M}\ket{k}
    =
    \sum_{k=1}^{M}\ket{k},
\end{equation}
and consequently
\begin{equation}
    \left(
        \mathbb I-\hat{\mathsf S}_{ij}
    \right)
    \ket{W_M}
    =
    0
\label{eq:supplement_w_swap_zero}
\end{equation}
for every edge $(i,j)\in\mathcal E_W$. Thus
\begin{equation}
    \hat H_W\ket{W_M}=0.
\label{eq:supplement_w_zero_energy}
\end{equation}

\subsubsection{Uniqueness}
\label{sec:supplement_w_uniqueness}

Let $\ket{\psi}$ be an arbitrary zero-energy state of
$\hat H_W$. The branch and defect constraints require
\begin{equation}
    \ket{\psi}
    =
    \sum_{k=1}^{M}
    c_k\ket{k}.
\label{eq:supplement_w_general_state}
\end{equation}
The coherent branch states $\{\ket{k}\}_{k=1}^{M}$ are linearly
independent for $\alpha\neq0$.

For every edge $(i,j)\in\mathcal E_W$, the zero-energy condition of
the swap penalty requires
\begin{equation}
    \hat{\mathsf S}_{ij}\ket{\psi}
    =
    \ket{\psi}.
\label{eq:supplement_w_swap_condition}
\end{equation}
Using Eq.~\eqref{eq:supplement_w_swap_branch_action}, the swapped
state is
\begin{equation}
    \hat{\mathsf S}_{ij}\ket{\psi}
    =
    c_i\ket{j}
    +
    c_j\ket{i}
    +
    \sum_{k\neq i,j}c_k\ket{k}.
\label{eq:supplement_w_swapped_general_state}
\end{equation}
Comparison with Eq.~\eqref{eq:supplement_w_general_state} gives
\begin{equation}
    c_i=c_j
\label{eq:supplement_w_equal_edge_coefficients}
\end{equation}
for every edge $(i,j)\in\mathcal E_W$.

Because the swap graph is connected, equality of coefficients along
every edge implies
\begin{equation}
    c_1=c_2=\cdots=c_M.
\label{eq:supplement_w_all_equal_coefficients}
\end{equation}
Therefore every zero-energy state is proportional to
\begin{equation}
    \sum_{j=1}^{M}\ket{j},
\end{equation}
and hence to $\ket{W_M}$. We conclude that
\begin{equation}
    \ker\hat H_W
    =
    \operatorname{span}
    \left\{
        \ket{W_M}
    \right\}.
\label{eq:supplement_w_unique_kernel}
\end{equation}
The exact constraint hierarchy is therefore
\begin{equation}
    2^M
    \longrightarrow
    M
    \longrightarrow
    1.
\label{eq:supplement_w_reduction}
\end{equation}

\section{Measurement of the bosonic swap constraint}
\label{sec:supplement_swap_measurement}

Although the bosonic swap operator
$\hat{\mathsf S}_{ij}$ is not a low-order polynomial in the local
oscillator operators, its expectation value can be measured through a
balanced mode transformation followed by a single-mode parity
measurement.

For the pair of modes $i$ and $j$, define the symmetric and
antisymmetric combinations
\begin{equation}
    \hat b_{ij,+}
    =
    \frac{
        \hat a_i+\hat a_j
    }{\sqrt{2}},
    \qquad
    \hat b_{ij,-}
    =
    \frac{
        \hat a_i-\hat a_j
    }{\sqrt{2}}.
\label{eq:supplement_symmetric_antisymmetric_modes}
\end{equation}
Under exchange of the two original modes,
\begin{equation}
    \hat{\mathsf S}_{ij}
    \hat a_i
    \hat{\mathsf S}_{ij}^{\dagger}
    =
    \hat a_j,
    \qquad
    \hat{\mathsf S}_{ij}
    \hat a_j
    \hat{\mathsf S}_{ij}^{\dagger}
    =
    \hat a_i.
\label{eq:supplement_swap_annihilation_action}
\end{equation}
Consequently,
\begin{equation}
    \hat{\mathsf S}_{ij}
    \hat b_{ij,+}
    \hat{\mathsf S}_{ij}^{\dagger}
    =
    \hat b_{ij,+},
    \qquad
    \hat{\mathsf S}_{ij}
    \hat b_{ij,-}
    \hat{\mathsf S}_{ij}^{\dagger}
    =
    -\hat b_{ij,-}.
\label{eq:supplement_swap_normal_mode_action}
\end{equation}
Thus the symmetric mode is invariant under the swap, whereas the
antisymmetric mode changes sign.

To make the corresponding operator identity explicit, consider the
two-mode Fock basis associated with the normal modes,
\begin{equation}
    \ket{n_{+},n_{-}}_{ij}
    =
    \frac{
        \left(
            \hat b_{ij,+}^{\dagger}
        \right)^{n_{+}}
        \left(
            \hat b_{ij,-}^{\dagger}
        \right)^{n_{-}}
    }{
        \sqrt{n_{+}!n_{-}!}
    }
    \ket{0,0}_{ij}.
\label{eq:supplement_normal_mode_fock_basis}
\end{equation}
Equation~\eqref{eq:supplement_swap_normal_mode_action} gives
\begin{equation}
    \hat{\mathsf S}_{ij}
    \ket{n_{+},n_{-}}_{ij}
    =
    (-1)^{n_{-}}
    \ket{n_{+},n_{-}}_{ij}.
\label{eq:supplement_swap_normal_mode_eigenvalue}
\end{equation}
The swap is therefore exactly equal to photon-number parity of the
antisymmetric mode,
\begin{equation}
    \boxed{
    \hat{\mathsf S}_{ij}
    =
    (-1)^{
        \hat b_{ij,-}^{\dagger}
        \hat b_{ij,-}
    }
    }
    \equiv
    \hat\Pi_{ij,-}.
\label{eq:supplement_swap_parity_identity}
\end{equation}
This identity holds on the complete two-mode Hilbert space and does
not rely on a large-$|\alpha|$ approximation or projection onto the
coherent-branch manifold.

\subsection{Balanced-beam-splitter measurement}
\label{sec:supplement_balanced_beam_splitter}

Let $\hat U_{ij}^{(\mathrm{BS})}$ denote a balanced beam-splitter
transformation chosen such that
\begin{equation}
    \left(
        \hat U_{ij}^{(\mathrm{BS})}
    \right)^{\dagger}
    \hat a_i
    \hat U_{ij}^{(\mathrm{BS})}
    =
    \hat b_{ij,+},
    \qquad
    \left(
        \hat U_{ij}^{(\mathrm{BS})}
    \right)^{\dagger}
    \hat a_j
    \hat U_{ij}^{(\mathrm{BS})}
    =
    \hat b_{ij,-}.
\label{eq:supplement_beam_splitter_transformation}
\end{equation}
The parity measured in the antisymmetric output port is then
\begin{align}
    \left(
        \hat U_{ij}^{(\mathrm{BS})}
    \right)^{\dagger}
    (-1)^{\hat a_j^{\dagger}\hat a_j}
    \hat U_{ij}^{(\mathrm{BS})}
    &=
    (-1)^{
        \hat b_{ij,-}^{\dagger}
        \hat b_{ij,-}
    }
    \nonumber\\
    &=
    \hat{\mathsf S}_{ij}.
\label{eq:supplement_beam_splitter_swap_identity}
\end{align}
For an arbitrary input state $\hat\rho$, the swap expectation value is
therefore
\begin{equation}
    \Tr
    \left(
        \hat\rho\hat{\mathsf S}_{ij}
    \right)
    =
    \Tr
    \left[
        \hat U_{ij}^{(\mathrm{BS})}
        \hat\rho
        \left(
            \hat U_{ij}^{(\mathrm{BS})}
        \right)^{\dagger}
        (-1)^{\hat a_j^{\dagger}\hat a_j}
    \right].
\label{eq:supplement_swap_measurement_expectation}
\end{equation}
Hence the measurement protocol consists of:

\begin{enumerate}
    \item mixing modes $i$ and $j$ on a balanced beam splitter;
    \item measuring the photon-number distribution in the
    antisymmetric output port;
    \item evaluating its parity.
\end{enumerate}

If $P_{ij,-}(n)$ is the measured photon-number probability in this
output mode, then
\begin{equation}
    \left\langle
        \hat{\mathsf S}_{ij}
    \right\rangle
    =
    \sum_{n=0}^{\infty}
    (-1)^n
    P_{ij,-}(n).
\label{eq:supplement_swap_from_photon_statistics}
\end{equation}
Accordingly, the expectation value of an individual swap penalty is
\begin{equation}
    \left\langle
        \mathbb I-\hat{\mathsf S}_{ij}
    \right\rangle
    =
    1-
    \left\langle
        \hat{\mathsf S}_{ij}
    \right\rangle,
\label{eq:supplement_swap_penalty_expectation}
\end{equation}
and the complete W-state symmetry contribution can be determined
edge by edge as
\begin{equation}
    \left\langle
        \hat H_{\mathrm{sym}}^{(W)}
    \right\rangle
    =
    \kappa
    \sum_{(i,j)\in\mathcal E_W}
    \left[
        1-
        \left\langle
            \hat{\mathsf S}_{ij}
        \right\rangle
    \right].
\label{eq:supplement_w_symmetry_measurement}
\end{equation}

This measurement relation does not imply that the interaction
$\mathbb I-\hat{\mathsf S}_{ij}$ is straightforward to realize as a
continuous physical Hamiltonian. Its implementation remains
platform-dependent. Nevertheless, the swap constraint is an exact
symmetry selector whose expectation value is directly accessible in
platforms supporting balanced mode mixing and photon-number parity
detection.

\section{Expectation values in target and orthogonal symmetry sectors}
\label{sec:supplement_expectation_comparison}

To illustrate the state selectivity of the parent Hamiltonians, we
compare their expectation values for the target states and for
symmetry-distinct states orthogonal to them. All coupling strengths
appearing in the corresponding selector terms are set to unity, and
no additional normalization of the Hamiltonians is introduced. The
resulting values therefore directly reflect the eigenvalues of the
violated positive-semidefinite constraints.

\subsection{GHZ parity sectors}
\label{sec:supplement_expectation_ghz}

Both $\ket{\mathrm{GHZ}_{+}}$ and
$\ket{\mathrm{GHZ}_{-}}$ satisfy the branch and alignment constraints,
so that
\begin{equation}
    \bra{\mathrm{GHZ}_{\pm}}
    \left(
        \hat H_{\mathrm{br}}
        +
        \hat H_{\mathrm{align}}^{(\mathrm{GHZ})}
    \right)
    \ket{\mathrm{GHZ}_{\pm}}
    =
    0.
\end{equation}
They differ only in their global-parity eigenvalues,
\begin{equation}
    \hat\Pi_G\ket{\mathrm{GHZ}_{\pm}}
    =
    \pm\ket{\mathrm{GHZ}_{\pm}}.
\end{equation}
Consequently, the parity Hamiltonian selecting the positive sector
gives
\begin{align}
    \bra{\mathrm{GHZ}_{+}}
    \hat H_{\mathrm{GHZ},+}
    \ket{\mathrm{GHZ}_{+}}
    &=0,
    \\
    \bra{\mathrm{GHZ}_{-}}
    \hat H_{\mathrm{GHZ},+}
    \ket{\mathrm{GHZ}_{-}}
    &=2\gamma.
\end{align}
Analogously,
\begin{align}
    \bra{\mathrm{GHZ}_{-}}
    \hat H_{\mathrm{GHZ},-}
    \ket{\mathrm{GHZ}_{-}}
    &=0,
    \\
    \bra{\mathrm{GHZ}_{+}}
    \hat H_{\mathrm{GHZ},-}
    \ket{\mathrm{GHZ}_{+}}
    &=2\gamma.
\end{align}
For $\gamma=1$, the wrong-parity state therefore has expectation
value $2$. The alignment coefficient does not contribute because both
GHZ states already belong to the aligned two-branch manifold.

\subsection{Cluster stabilizer sectors}
\label{sec:supplement_expectation_cluster}

Inside the four-dimensional pair-selected manifold, let
$\ket{C_{pq}}$, with $p,q\in\{0,1\}$, denote the simultaneous
eigenstates of the two graph-correlation operators,
\begin{equation}
    \hat K_A\ket{C_{pq}}
    =
    (-1)^p\ket{C_{pq}},
    \qquad
    \hat K_B\ket{C_{pq}}
    =
    (-1)^q\ket{C_{pq}}.
\label{eq:supplement_cluster_sector_definition}
\end{equation}
The target is $\ket{C_{00}}$. All four states satisfy the branch and
pair-alignment constraints, and hence only
$\hat H_{\mathrm{corr}}^{(C)}$ contributes to their energies. Using
\begin{equation}
    \hat H_{\mathrm{corr}}^{(C)}
    =
    \mu_A
    \left(
        \mathbb I-\hat K_A
    \right)^\dagger
    \left(
        \mathbb I-\hat K_A
    \right)
    +
    \mu_B
    \left(
        \mathbb I-\hat K_B
    \right)^\dagger
    \left(
        \mathbb I-\hat K_B
    \right),
\end{equation}
one obtains
\begin{align}
    \bra{C_{pq}}
    \hat H_C
    \ket{C_{pq}}
    ={}&
    \mu_A
    \left|
        1-(-1)^p
    \right|^2
    \nonumber\\
    &+
    \mu_B
    \left|
        1-(-1)^q
    \right|^2.
\label{eq:supplement_cluster_sector_energies}
\end{align}
Thus, for $\mu_A=\mu_B=1$,
\begin{equation}
    E_{00}=0,
    \qquad
    E_{10}=4,
    \qquad
    E_{01}=4,
    \qquad
    E_{11}=8.
\label{eq:supplement_cluster_numerical_energies}
\end{equation}
The factor of four associated with each violated stabilizer follows
from the squared penalty,
\begin{equation}
    \left|
        1-(-1)
    \right|^2
    =
    4.
\end{equation}
The state $\ket{C_{11}}$ violates both graph constraints and therefore
has twice the energy of $\ket{C_{10}}$ and $\ket{C_{01}}$. The
pair-alignment coefficient does not contribute because all four
comparison states lie in the pair-selected manifold.

\subsection{W permutation-symmetry sectors}
\label{sec:supplement_expectation_w}

For the three-mode case, let
\begin{equation}
    \omega
    =
    e^{2\pi i/3}
\end{equation}
and define the three Fourier combinations of the one-positive-branch
configurations,
\begin{equation}
    \ket{W_q}
    =
    \mathcal N_q
    \sum_{j=1}^{3}
    \omega^{q(j-1)}
    \ket{j},
    \qquad
    q=0,1,2.
\label{eq:supplement_w_fourier_states}
\end{equation}
The state $\ket{W_0}$ is the permutation-symmetric target, whereas
$\ket{W_1}$ and $\ket{W_2}$ are orthogonal symmetry-sector states.
All three satisfy the branch and fixed-defect constraints.

For the connected chain
\begin{equation}
    \mathcal E_W
    =
    \left\{
        (1,2),(2,3)
    \right\},
\end{equation}
the symmetry Hamiltonian is
\begin{equation}
    \hat H_{\mathrm{sym}}^{(W)}
    =
    \kappa
    \left[
        \left(
            \mathbb I-\hat{\mathsf S}_{12}
        \right)
        +
        \left(
            \mathbb I-\hat{\mathsf S}_{23}
        \right)
    \right].
\end{equation}
The symmetric target satisfies
\begin{equation}
    \left\langle
        \hat{\mathsf S}_{12}
    \right\rangle_{W_0}
    =
    \left\langle
        \hat{\mathsf S}_{23}
    \right\rangle_{W_0}
    =
    1,
\end{equation}
and therefore
\begin{equation}
    \bra{W_0}\hat H_W\ket{W_0}=0.
\end{equation}
For either orthogonal Fourier state,
\begin{equation}
    \left\langle
        \hat{\mathsf S}_{12}
    \right\rangle_{W_q}
    =
    \left\langle
        \hat{\mathsf S}_{23}
    \right\rangle_{W_q}
    =
    0,
    \qquad
    q=1,2.
\end{equation}
It follows that
\begin{equation}
    \bra{W_q}\hat H_W\ket{W_q}
    =
    2\kappa,
    \qquad
    q=1,2.
\label{eq:supplement_w_sector_energies}
\end{equation}
For $\kappa=1$, both orthogonal W-sector states therefore have
expectation value $2$. The fixed-defect coefficient does not
contribute because all three states remain inside the one-positive-
branch manifold.

\subsection{Numerical comparison}
\label{sec:supplement_expectation_figure}

\begin{figure*}[h]
    \centering
    \includegraphics[
        width=\textwidth
    ]{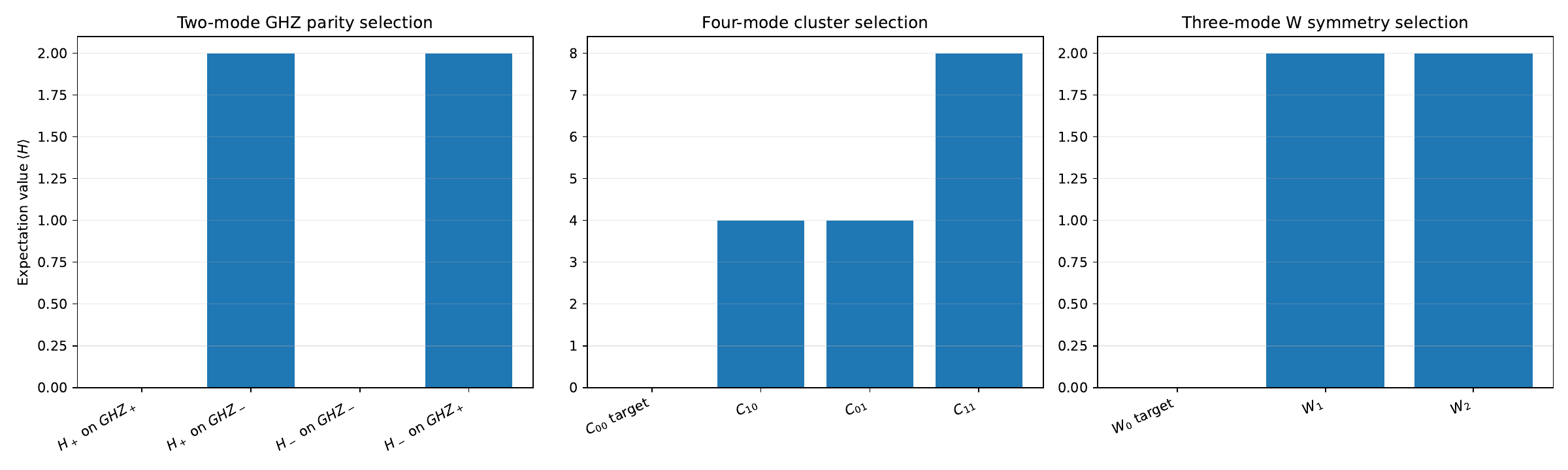}
    \caption{
    Parent-Hamiltonian expectation values for the target states and
    orthogonal symmetry-sector states. Left: the GHZ state selected by
    the chosen global parity has zero energy, while the opposite-parity
    state has energy $2$ for $\gamma=1$. Center: the cluster target
    $\ket{C_{00}}$ has zero energy; violating one squared graph
    constraint gives energy $4$, while violating both gives energy $8$
    for $\mu_A=\mu_B=1$. Right: the symmetric three-mode W state has
    zero energy, whereas the two orthogonal Fourier-sector states have
    energy $2$ for a two-edge swap graph with $\kappa=1$. All branch
    and intermediate selection terms vanish on the displayed states,
    and no normalization of the Hamiltonians has been introduced.
    }
    \label{fig:supplement_expectation_comparison}
\end{figure*}

Figure~\ref{fig:supplement_expectation_comparison} summarizes these
expectation values. The target state in each example has zero energy,
whereas states violating the final parity, stabilizer, or permutation
constraint acquire a finite value. The different numerical scales
arise from the definitions of the selector terms: the GHZ and W
penalties are linear in $\mathbb I-\hat U$, while the cluster
constraints use the squared form
$(\mathbb I-\hat K)^\dagger(\mathbb I-\hat K)$. No rescaling has been
applied.

\end{document}